\documentclass[aps,pra,twocolumn,superscriptaddress,showpacs,showkeys,amsmath,amssymb]{revtex4}

\usepackage{amsfonts}
\usepackage{amssymb,amsmath}
\usepackage{mathrsfs}
\usepackage{latexsym}
\usepackage{amsmath}
\usepackage[cp1251]{inputenc}
\usepackage{graphicx}
\usepackage{dcolumn}
\usepackage{bm}
\usepackage{color}

\begin{document}

    \title{Polaron in dilute 2D Bose gas at low temperatures}

    \author{Volodymyr~Pastukhov\footnote{e-mail: volodyapastukhov@gmail.com}}
    \affiliation{Department for Theoretical Physics, Ivan Franko National University of Lviv,\\ 12 Drahomanov Street, Lviv-5, 79005, Ukraine}




    \begin{abstract}
  The properties of a Bose polaron immersed in a dilute two-dimensional medium at finite temperatures are discussed. Assuming that the impurity is weakly-coupled to the bath particles we have perturbatively calculated the polaron energy, effective mass, quasiparticle residue and damping rate. The parameters of impurity spectrum are found to be well-defined in the whole temperature region whereas the pole structure of the impurity Green's function is visible only at absolute zero. At any finite temperatures the quasiparticle residue is logarithmically divergent signalling of the branch-cut behavior of the polaron propagator.
    \end{abstract}

    \maketitle

\section{Introduction}
\label{sec1}
\setcounter{equation}{0}
It is well-know that the physics is highly non-trivial in low dimensions, where the presence of gapless excitations in the energy spectrum usually leads to the dramatic change of system's properties, particularly to the emergence of an off-diagonal long-range order in one dimension at absolute zero and in the two-dimensional case at low but finite temperatures. In this context a very important issue is the behavior of low-dimensional impurity particles immersed in the bosonic baths \cite{Astrakharchik_Pitaevskii,Parisi,Loft,Grusdt_Astrakharchik}. Recently the three-dimensional counterpart of this problem was extensively discussed theoretically by means of diagrammatic \cite{Novikov_Ovchinnikov_09,Rath,Christensen,Panochko_et_al}, variational \cite{Tempere_et_al,Novikov_Ovchinnikov_10,Li,Shchadilova}, renormalization group \cite{Grusdt_et_al_1,Grusdt_et_al_2} and numerical \cite{Vlietinck,Ardila_1,Ardila_2} methods. Besides experimental realization \cite{Catani_et_al} the low-dimensional Bose polarons are interesting from the methodological point of view, where the conventional perturbative approaches generally break due to the long-range nature of the boson-impurity effective interaction. But the great advantage of one-dimensional impurity problem is the existence of an exact solution \cite{McGuire_65,McGuire_66,Castella_Zotos,Gamayun} in the equal-mass limit which could serve a benchmark for any approximate calculation schemes. The properties of impurity particles placed in two-dimensional Bose system are less studied \cite{Dehkharghani,Grusdt,Grusdt_Fleischhauer}, especially in the finite-temperature region. Actually at present time little is known about the full temperature dependence of the key parameters of the Bose polaron spectrum even in three dimensions and therefore the problem is of current interest \cite{Guenther,Levinsen_et_al,Bosepolaron_D}. Trying to fill this gap we have considered in present paper the leading-order low-temperature properties of a polaron weakly-coupled to the dilute two-dimensional Bose gas.

\vspace*{-0.45cm} 
\section{Formulation}
The Euclidean action of the Bose-Fermi mixture immersed in a two-dimensional volume $A$ at temperature $T$ reads
\begin{eqnarray}\label{S}
	S=S_0+S_B+S_{int},
\end{eqnarray}
where the first term describes gas of spinless non-interacting fermions of mass $m_i$ with chemical potential $\mu_i$
\begin{eqnarray}\label{S_0}
	S_0=\int dx\, \psi^*(x)\left\{\partial_{\tau}+\frac{\hbar^2\nabla^2}{2m_i}+\mu_i\right\}\psi(x),
\end{eqnarray}
(here $x=(\tau, {\bf r})$ and $\int dx=\int_0^{1/T}d\tau \int_A d{\bf r}$); the second one refers to the Bose condensate states
\begin{eqnarray}\label{S_B}
	S_B=\int dx\, \phi^*(x)\left\{\partial_{\tau}+\frac{\hbar^2\nabla^2}{2m}+\mu\right\}\phi(x)\nonumber\\
	-\frac{g}{2}\int dx |\phi(x)|^4.
\end{eqnarray}
The last term of (\ref{S}) 
\begin{eqnarray}\label{S_int}
S_{int}=-\tilde{g}\int dx\, |\phi(x)|^2\psi^*(x)\psi(x),
\end{eqnarray}
takes into account the interaction between Bose and Fermi constituents. The path integral is carried out over antiperiodic in imaginary-time variable $\tau$ Grassmann fields $\psi^*(x)$, $\psi(x)$ and periodic complex field $\phi(x)$. The bare coupling constants $g$ and $\tilde{g}$ are generally dependent on the short-range behavior of the two-body potentials and will be specified below. Assuming that both the boson-boson and the boson-impurity interactions are weak we apply the Popov prescription \cite{Popov} separating  ``slowly'' and ``rapidly'' varying modes $\psi(x)=\psi_{<}(x)+\psi_{>}(x)$, $\phi(x)=\phi_{<}(x)+\phi_{>}(x)$, where $\psi_{>}(x)$, $\phi_{>}(x)$ contain Fourier harmonics with $|{\bf k}|>\Lambda$. The introduced here auxiliary the inverse length-scale parameter $\Lambda$ in the thermodynamic limit is fully determined by the properties of Bose subsystem and for the dilute gas $\Lambda^2$ is of order density $n$ of the system. Then, choosing arbitrary short-ranged two-body potentials (Gaussian in our case) and integrating out the ``rapidly'' varying fields with the additional assumption that the characteristic quantity with dimension of energy $\hbar^2\Lambda^2/m(m_i)$ is much larger that other energy scales, namely the chemical potential of Bose gas, the biding energy of an impurity and temperature one obtains the effective action $S_{eff}$ governing the properties of ``slowly'' varying fields $\phi_{<}(x)$, $\psi_{<}(x)$. Actually this action is identical \cite{Konietin} to (\ref{S}) but with bare coupling constants $g$ and $\tilde{g}$ replaced by the $t$-matrices $t=2\pi \hbar^2/\left(m\ln\left[2e^{-\gamma}/a\Lambda\right]\right)$ and $\tilde{t}=\pi \hbar^2(m+m_i)/\left(mm_i\ln\left[2e^{-\gamma}/\tilde{a}\Lambda\right]\right)$ ($\gamma=0.57721\ldots$ is the Euler-Mascheroni constant), respectively which in turn are characterized by $s$-wave scattering lengths $a$ and $\tilde{a}$. Adopting the phase-density representation for the bosonic fields $\phi^*(x)=\sqrt{n(x)}e^{-i\varphi(x)}$, $\phi(x)=\sqrt{n(x)}e^{i\varphi(x)}$ (from now on the subscript ${<}$ is omitted) we proceed with the hydrodynamic approach, which correctly determines the low-temperature properties of dilute Bose systems in any dimension. After all these transformations the effective action reads
\begin{align}\label{S_eff}
S_{eff}=\int dx\, \left\{n(x)i\partial_{\tau}\varphi(x)-\mu n(x)-\frac{1}{2}tn^2(x)\right.\nonumber\\
\left.-\frac{\hbar^2}{2m}n(x)[\nabla\varphi(x)]^2-\frac{\hbar^2}{8m}\frac{[\nabla n(x)]^2}{n(x)}\right\}\nonumber\\
+\int dx\, \psi^*(x)\left\{\partial_{\tau}+\hbar^2\nabla^2/2m_i+\mu_i\right\}
\psi(x)\nonumber\\
-\tilde{t}\int dx\,n(x)\psi^*(x)\psi(x).
\end{align}
In our formulation, which is accurate in the extremely dilute limit at low temperatures the thermodynamic relation $nA=-\partial \Omega/\partial \mu$, applied to the system with continuous translational symmetry, identifies \cite{Pastukhov_InfraredStr} the zero-momentum Fourier transform of $n(x)$ with the density of a Bose gas. However, the inclusion of external potential \cite{Johnson} or even account of the finiteness of particle number \cite{Sacha,Volosniev} in the bath system can drastically change this result. The hydrodynamic approach that is free of infrared divergences in the perturbation theory automatically guarantees for the excitation spectrum of Bose system to be gapless in the ordered phase. Furthermore, the long-length parameters of spectrum can be related (see Ref.~\cite{Pastukhov_16} for a simple derivation) to the macroscopic observables, namely inverse compressibility and superfluid density of the system. Another important feature of this formulation is the presence of off-diagonal long-range order with the exponent which is characteristic for the Berezinskii–Kosterlitz–Thouless (BKT) transition.

Expanding $S_{eff}$ in terms of phase and density fluctuations up to the quadratic terms, making use of the Fourier transformation and change of variables in the path-integral
$b_{K}=i\sqrt{n/\alpha_{k}}\,\varphi_{
	K}+\frac{1}{2}\sqrt{\alpha_{k}/n}\,n_K$,
$b^*_{K}=-i\sqrt{n/\alpha_{k}}\,\varphi_{-
	K}+\frac{1}{2}\sqrt{\alpha_{k}/n}\,n_{-K}$, where $\alpha_k=E_k/\varepsilon_k$ ($\varepsilon_k=\hbar^2k^2/2m$, $E_k=\sqrt{\varepsilon^2_k+2nt\varepsilon_k}$ is the Bogoliubov spectrum), $\varphi_K$ and $n_K$ are $(2+1)$ Fourier transforms of $\varphi(x)$ and $n(x)$, respectively we finally obtain
\begin{align}\label{S_simple}
S_{eff}=\sum_K\left\{i\omega_k-E_k\right\}b^*_Kb_K+\sum_P\left\{i\nu_p-\xi_p\right\}\psi^*_P\psi_P\nonumber\\
 -\sqrt{\frac{T}{A}}\sum_{K,P}\tilde{t}\sqrt{\frac{n}{\alpha_{k}}}\left\{b^*_K+b_{-K}\right\}\psi^*_P\psi_{P+K},
\end{align} 
here $\omega_k, \nu_p$ are bosonic and fermionic Matsubara frequencies and $\xi_p=\hbar^2k^2/2m_i-\tilde{\mu}_i$ denotes impurity spectrum with the mean-field shifted chemical potential $\tilde{\mu}_i=\mu_i-n\tilde{t}$.
Equation (\ref{S_simple}) is nothing but the action of the Fr\"ohlich model which is known to be adequate for a weakly-coupled Bose polaron only. The extension on stronger interactions necessarily requires the inclusion not only the emission and absorption of a single phonon by the impurity particle but also two-phonon scattering processes \cite{Grusdt_et_al_2}. In order to reach the strong-coupling Bose polaron limit in the hydrodynamic formulation one has to go beyond the action (\ref{S_simple}) and take into account the anharmonic terms which are responsible for the interaction of the Bogoliubov quasiparticles. However, assuming weakness of the boson-impurity interaction and low temperatures we strongly restrict the average number of excited phonons in the system which allows to describe the properties of two-dimensional Bose polaron within the Fr\"ohlich model.

To find out the impact of interaction on the impurity properties we calculate the single-particle Green's function $G(P)=\langle\psi^*_P\psi_P\rangle$ (here $\langle\ldots\rangle$ denotes statistical averaging with action (\ref{S_simple}))
\begin{align}\label{G}
G^{-1}(P)=i\nu_p-\xi_p-\Sigma(P),
\end{align} 
where the self-energy on the one-loop level is plotted in Fig.~1.
\begin{figure}
	\centerline{\includegraphics
		[width=0.45\textwidth,clip,angle=-0]{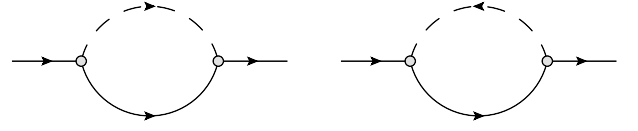}}
	\caption{Diagrams contributing to the self-energy on the one-loop level. Solid and dashed lines correspond to fermionic and bosonic propagators, respectively. The boson-impurity bare vertex $\tilde{t}\sqrt{n/\alpha_{k}}$ is denoted by dot.}
\end{figure}
The explicit evaluation in the limit of vanishing impurity density yields
\begin{align}\label{Sigma}
\Sigma(P)=-\frac{1}{A}\sum_{|{\bf k}|<\Lambda}\frac{n\tilde{t}^2}{\alpha_k}\left\{
\frac{1+n(E_k/T)}{E_k+\xi_{|{\bf k}+{\bf p}|}-i\nu_p}\right.\nonumber\\
\left.-\frac{n(E_k/T)}{E_k-\xi_{|{\bf k}+{\bf p}|}+i\nu_p}\right\},
\end{align} 
(where $n(x)=1/[e^x-1]$ is the Bose distribution) which after analytical continuation $\Sigma(P)_{\nu_p\to \nu+i0}=\Sigma_R(\nu,p)+i\Sigma_I(\nu,p)$ can be used for calculations of the renormalized impurity spectrum
\begin{align}\label{xi*}
\xi^*_p=\xi_p+\Sigma_R(\xi_p,p),
\end{align} 
damping
\begin{align}\label{Gamma}
\Gamma_p=-\Sigma_I(\xi_p,p),
\end{align} 
and quasiparticle residue
\begin{align}\label{Z}
Z^{-1}_p=1-\frac{\partial \Sigma_R(\xi_p,p)}{\partial \xi_p}.
\end{align}

\section{Results and discussion}
It is naturally to start our analysis of the polaron properties from the low-temperature region, where the leading-order corrections (note that $a^2n,\tilde{a}^2n\ll 1$) to the impurity binding energy
\begin{align}\label{xi*0}
\xi^*_0=\frac{2\pi\hbar^2n(1+\alpha)}{m\alpha|\ln[\tilde{a}^2n]|}\left\{1-\frac{\ln|\ln[a^2n]|}{|\ln[\tilde{a}^2n]|}\right.\nonumber\\
\left.+\frac{2\gamma+\ln\pi+\frac{2\alpha}{\alpha-1}\ln\left[\frac{2\alpha}{\alpha+1}\right]}
{|\ln[\tilde{a}^2n]|}\right\},
\end{align}
($\alpha=m_i/m$) polaron effective mass
\begin{eqnarray}
\frac{m_i}{m^*_i}=1-\frac{|\ln[a^2n]|}{\ln^2[\tilde{a}^2n]}\frac{1}{\alpha},
\end{eqnarray}
and wave-function renormalization
\begin{eqnarray}
Z^{-1}_0=1+\frac{|\ln[a^2n]|}{\ln^2[\tilde{a}^2n]}\frac{1+\alpha}{2\alpha},
\end{eqnarray}
can be calculated analytically in the long-length limit. At zero-temperature the motionless polaron does not loose its energy by producing phonons and that is why the damping is absent $\Gamma_0=0$ in this case. The adopted approximation is insensible to the quantum statistical effects and therefore it is not surprising that formula (\ref{xi*0}) in the equal-mass limit (and with $\tilde{a}\to a$) coincides to the leading order with the result of Ref.~\cite{Astrakharchik_etal} for the chemical potential of a two-dimensional Bose gas.

The explicit result (\ref{Sigma}) for the one-loop polaron self-energy indicates that the impurity spectrum is well-defined at finite temperatures even in one dimension. It also should be noted that the above-used approximation is valid only in the low-temperature region far from the BKT transition. The first estimation for the critical temperature of  dilute two-dimensional Bose gas was given long ago by Popov $T_{BKT}\simeq \frac{2\pi\hbar^2 n}{m\ln[C\hbar^2/mt]}$. The precise Monte Carlo simulations \cite{Prokofev} of the classical $|\psi|^4$-model generally confirmed this behavior but the value of constant was found to be large $C=380\pm3$. The latter greatly restricts the temperature region where our results are applicable. Furthermore, the Bogoliubov approximation used in the present study is accurate only when the normal density of superfluid (which is of order $T^3$ at low temperatures in 2D) is small. These all limit us to consider in the following only a narrow region between absolute zero and $T\sim nt$. 

Making use of notations for the dimensionless shift of binding energy
\begin{align}
\Sigma_R(-\tilde{\mu}_i,0)-\Sigma_R(-\tilde{\mu}_i,0)|_{T=0}=
\frac{n\tilde{t}}{|\ln[\tilde{a}^2n]|}\epsilon(\alpha,\tilde{T}),
\end{align}
(where $\tilde{T}=T/nt$), effective mass
\begin{eqnarray}
\frac{m_i}{m^*_i}=1-\frac{|\ln[a^2n]|}{\ln^2[\tilde{a}^2n]}\Delta(\alpha,\tilde{T}),
\end{eqnarray}
and damping rate ($\alpha< 1$)
\begin{eqnarray}
\frac{\Gamma_0}{n\tilde{t}}=\frac{\pi}{|\ln[\tilde{a}^2n]|}\frac{\alpha}{1-\alpha}
n\left(2\alpha/[(1-\alpha^2)\tilde{T}]\right),
\end{eqnarray}
which is calculated analytically to the very end, we obtain the finite-temperature behavior of the Bose polaron spectrum at small momenta. The typical behavior of functions 
$\epsilon(\alpha,\tilde{T})$, $\Delta(\alpha,\tilde{T})$ is presented in Figs.~2-3.
\begin{figure}
	\centerline{\includegraphics
		[width=0.55\textwidth,clip,angle=-0]{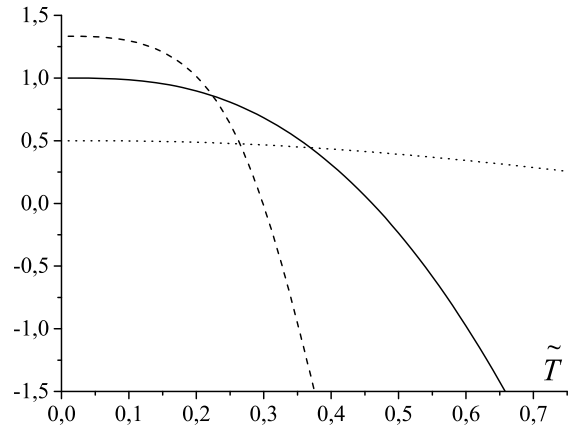}}
	\caption{Functions $\Delta(3/4,\tilde{T})$ (dashed), $\Delta(1,\tilde{T})$ (solid) and $\Delta(2,\tilde{T})$ (dotted) determining temperature dependence of the effective mass.}
\end{figure}
\begin{figure}
	\centerline{\includegraphics
		[width=0.55\textwidth,clip,angle=-0]{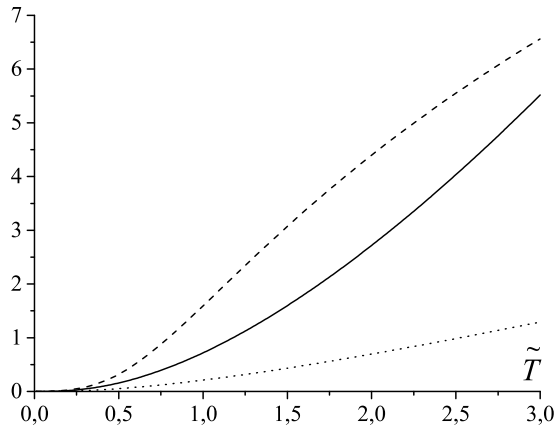}}
	\caption{Dimensionless finite-temperature binding energy $\epsilon(3/4,\tilde{T})$ (dashed), $\epsilon(1,\tilde{T})$ (solid) and $\epsilon(2,\tilde{T})$ (dotted).}
\end{figure}

The most unexpected peculiarity of the considered system is met during the calculations of quasiparticle residue at finite temperatures. In particular, we have found that $Z^{-1}_p$ is divergent at {\it any} momenta. The logarithmic character of these divergences is the same for every wave-vector ${\bf p}$ which leads us to conclusion that behavior of the retarded polaron Green's function near singular point is the following
\begin{eqnarray}\label{Gsing}
\lim_{p \to 0}G_{ret}(\nu,p)|_{\nu\to \xi^*_p}\propto \frac{1}{[\nu-\xi^*_p]^{1-\eta}}.
\end{eqnarray}
In general, the exponent $\eta$ is momentum-dependent with the value 
\begin{eqnarray}
\eta^{(1)}=\frac{mT}{2\pi\hbar^2n}\frac{\tilde{t}^2}{t^2},
\end{eqnarray} 
calculated on the one-loop level in the long-length limit. Actually this formula is nothing but the low-temperature leading-order result for the exponent $\eta$. The same structure of the Green's function is intrinsic for the 1D Bose polarons \cite{Pastukhov_1D} at $T=0$. The formulas for $\eta$s are different in these two cases but the reasons for appearing of such a non-analytical behavior are the same. Below the transition temperature $T_{BKT}$, i.e., in an ordered phase the density fluctuations of Bose gas are strongly developed (the appropriate situation is also observed in 1D at $T=0$) and consequently the effective interaction potential between Bose polaron and bath particles is long-ranged (even if the bare one is short-ranged) which causes such a power-law decay of the impurity one-body density matrix at large distances. A similar logarithmically-divergent behavior of the quasiparticle residue and effective mass is intrinsic and for the $D$-dimensional systems \cite{Bosepolaron_D} at the Bose-Einstein condensation point and for the two-dimensional Bose polaron interacting with the Tkachenko modes \cite{Caracanhas}.

\section{Conclusions}
In summary, we have studied the properties of the single impurity particle immersed in the two-dimensional dilute Bose bath. It was shown by means of the one-loop perturbative calculations that the spectrum of a Bose polaron is well-defined at temperatures below the Berezinskii–Kosterlitz–Thouless transition point while the quasiparticle residue is always logarithmically divergent. The presence of these divergences is treated as a non-pole behavior of the impurity Green's function with the non-universal exponent which was evaluated in the first order of perturbation theory.

\begin{center}
	{\bf Acknowledgements}
\end{center}
We thank Dr.~A.~Rovenchak for fruitful discussions.
This work was partly supported by Project FF-30F (No.~0116U001539) from the Ministry of Education and Science of Ukraine.

\end{document}